\documentclass[reprint,amsmath,amssymb, aps,pre,floatfix,superscriptaddress]{revtex4-2}

\usepackage{graphicx} 
\usepackage{siunitx}
\usepackage[version=4]{mhchem}
\usepackage{comment}
\usepackage{breqn}
\usepackage{xcolor}
\usepackage{tikz}
\usepackage{scalerel}
\usepackage{xifthen}
\usepackage{chngcntr}
\usepackage{xcolor}
\usepackage{bm}
\usepackage{physics}
\usepackage{tcolorbox}
\usepackage[utf8]{inputenc}
\usepackage[T1]{fontenc}
\usetikzlibrary{svg.path}
\usepackage[colorlinks,allcolors=blue]{hyperref}

\definecolor{orcidlogocol}{HTML}{A6CE39}
\tikzset{
  orcidlogo/.pic={
    \fill[orcidlogocol] svg{M256,128c0,70.7-57.3,128-128,128C57.3,256,0,198.7,0,128C0,57.3,57.3,0,128,0C198.7,0,256,57.3,256,128z};
    \fill[white] svg{M86.3,186.2H70.9V79.1h15.4v48.4V186.2z}
 svg{M108.9,79.1h41.6c39.6,0,57,28.3,57,53.6c0,27.5-21.5,53.6-56.8,53.6h-41.8V79.1z M124.3,172.4h24.5c34.9,0,42.9-26.5,42.9-39.7c0-21.5-13.7-39.7-43.7-39.7h-23.7V172.4z}
 svg{M88.7,56.8c0,5.5-4.5,10.1-10.1,10.1c-5.6,0-10.1-4.6-10.1-10.1c0-5.6,4.5-10.1,10.1-10.1C84.2,46.7,88.7,51.3,88.7,56.8z};
  }
}

\newcommand\orcid[1]{\href{https://orcid.org/#1}{\mbox{\scalerel*{
\begin{tikzpicture}[yscale=-1,transform shape]
\pic{orcidlogo};
\end{tikzpicture}
}{|}}}}

\begin{document}

\setlength{\unitlength}{1cm}

\def \TITLE {Using large language models for parametric shape optimization}

\def\ADD#1{{\textcolor{magenta}{#1}}}

\def \SI {SI}
\def \viz {\textit{viz.},~}
\def \ie {\textit{i.e.},~}
\def \eg {\textit{e.g.},~}

\def \dofone {d_1}
\def \doftwo {d_2}
\def \dofc {d_{\text{c}}}
\def \dotdofone {\dot{d_1}}
\def \dotdoftwo {\dot{d_2}}
\def \dotdofone {\dot{d}_1}
\def \dotdoftwo {\dot{d}_2}
\def \dotdofc {\dot{d}_{\text{c}}}

\def \dofmin {d_{\text{min}}}
\def \dofmax {d_{\text{max}}}

\def \brc {\br_{\text{c}}}
\def \bri {\br_{\text{i}}}

\def \brcx {\br_{\text{cx}}}
\def \brcy {\br_{\text{cy}}}
\def \dotbrc {\dot{\br}_{\text{c}}}
\def \dotbrc {\dot{\br}_{\text{c}}}
\def \dotbri {\dot{\br}_{\text{i}}}
\def \u {u}
\def \v {v}
\def \bFi {\bF_i}
\def \bFj {\bF_j}
\def \bFk {\bF_k}

\def \bex {\be_x}
\def \bey {\be_y}
\def \bez {\be_z}

\def \mX {\langle X \rangle}

\def \ns {n_{\text{ht}}}
\def \Re {\mathrm{Re}}

\def \Dr {D_r}
\def \al {LLM-PSO}
\def \iter {n}
\def \bxopt {\bx^{\ast}}
\def \De {\text{De}}
\def \Re {\text{Re}}
\def \Ma {\text{Ma}}
\def \Rt {\mR}

\def \mA {\mathcal{A}}
\def \mD {\mathcal{D}}
\def \mC {\mathcal{C}}

\def \be {\mathbf{e}}
\def \bf {\mathbf{f}}
\def \bq {\mathbf{q}}
\def \br {\mathbf{r}}
\def \bt {\mathbf{t}}

\def \bu {\mathbf{u}}
\def \bv {\mathbf{v}}
\def \bx {\mathbf{x}}
\def \bw {\mathbf{w}}

\def \bsigma {\boldsymbol{\sigma}}
\def \btau {\boldsymbol{\tau}}

\def \bzero {\boldsymbol{0}}

\def \bA {\mathbf{A}}
\def \bB {\mathbf{B}}
\def \bC {\mathbf{C}}
\def \bD {\mathbf{D}}
\def \bE {\mathbf{E}}
\def \bF {\mathbf{F}}
\def \bG {\mathbf{G}}
\def \bH {\mathbf{H}}
\def \bI {\mathbf{I}}
\def \bJ {\mathbf{J}}
\def \bK {\mathbf{K}}
\def \bL {\mathbf{L}}
\def \bM {\mathbf{M}}
\def \bN {\mathbf{N}}
\def \bO {\mathbf{O}}
\def \bP {\mathbf{P}}
\def \bQ {\mathbf{Q}}
\def \bR {\mathbf{R}}
\def \bS {\mathbf{S}}
\def \bT {\mathbf{T}}
\def \bU {\mathbf{U}}
\def \bV {\mathbf{V}}

\def \bOmega {\boldsymbol{\Omega}}
\def \bomega {\boldsymbol{\omega}}
\def \bell {\boldsymbol{\ell}}

\def \bGamma {\boldsymbol{\Gamma}}

\def \bn {\mathbf{n}}
\def \bI {\mathbf{I}}

\def \tbu {\tilde{\bu}}
\def \tbr {\tilde{\br}}
\def \tbR {\tilde{\bR}}
\def \tbU {\tilde{\bU}}
\def \tbE {\tilde{\bE}}
\def \tbF {\tilde{\bF}}

\def \tbOmega {\tilde{\bOmega}}
\def \tbGamma {\tilde{\bGamma}}
\def \tbtau {\tilde{\btau}}
\def \tbsigma {\tilde{\bsigma}}

\def \ta {\tilde{a}}
\def \tc {\tilde{c}}

\def \tp {\tilde{p}}
\def \ts {\tilde{s}}

\def \tt {\tilde{t}}
\def \tx {\tilde{x}}
\def \ty {\tilde{y}}

\def \tF {\tilde{F}}
\def \tG {\tilde{G}}

\def \tU {\tilde{U}}
\def \tV {\tilde{V}}

\def \tGamma {\tilde{\Gamma}}
\def \tOmega {\tilde{\Omega}}

\def \tgrad {\tilde{\grad}}

\def \mF {\mathcal{F}}
\def \mI {\mathcal{I}}
\def \mM {\mathcal{M}}
\def \mN {\mathcal{N}}

\def \mR {\mathcal{R}}
\def \mV {\mathcal{V}}
\def \mO {\mathcal{O}}

\def \lp {\left(}
\def \rp {\right)}

\def \ls {\left[}
\def \rs {\right]}

\def \d {\text{d}}
\def \dr {\d r}

\def \tran {\mathsf{T}}
\def \nfree {n_{\text{F}}}
\def \nini {n_{\text{ini}}}

\newcommand{\nus}{Department of Mechanical Engineering, National University of Singapore, 117575, Singapore}

\title{\TITLE}
\author{Xinxin Zhang~\orcid{0000-0003-0297-7061}}
\affiliation{\nus}
\author{Zhuoqun Xu~\orcid{0000-0002-3535-4402}}
\affiliation{\nus}
\author{Guangpu Zhu~\orcid{0000-0002-7721-0685}}
\affiliation{\nus}
\author{Chien Ming Jonathan Tay~\orcid{0000-0002-3759-5146}}
\affiliation{\nus}
\author{Yongdong Cui~\orcid{0000-0003-3052-5337}}
\affiliation{\nus}
\author{Boo Cheong Khoo~\orcid{0000-0003-4710-4598}}
\affiliation{\nus}
\author{Lailai Zhu~\orcid{0000-0002-3443-0709}}
\email{lailai\_zhu@nus.edu.sg}
\affiliation{\nus}
\date{\today}

\begin{abstract}
Recent advanced large language models (LLMs) have showcased their emergent capability of in-context learning, facilitating intelligent decision-making through natural language prompts without retraining. 
This new machine learning paradigm has shown promise in various fields, including general control and optimization problems. Inspired by these advancements, we explore the potential of LLMs for a specific and essential engineering task: parametric shape optimization (PSO).
We develop an optimization framework, \al, that leverages an LLM to determine the optimal shape of parameterized engineering designs in the spirit of evolutionary strategies. 
Utilizing the ``Claude 3.5 Sonnet'' LLM,  we evaluate \al ~on two benchmark flow optimization problems, specifically aiming to identify drag-minimizing profiles for 1) a two-dimensional airfoil in laminar flow, and 2) a three-dimensional axisymmetric body in Stokes flow.
In both cases,  \al ~successfully identifies optimal shapes in agreement with benchmark solutions. Besides, it generally converges faster than other classical optimization algorithms. Our preliminary exploration may inspire further investigations into harnessing LLMs for shape optimization and engineering design more broadly.
\end{abstract}

\maketitle

\section{Introduction}

Shape optimization is a field of mathematical and computational design aimed at determining the best geometric configuration of a domain to optimize a given objective while satisfying constraints. It finds applications across engineering and science, from aerodynamic design~\cite{li2020efficient} to structural analysis~\cite{zong2018two,gong2023shape}, where efficient and innovative shapes are crucial for performance enhancement. By leveraging tools such as numerical simulations, optimization algorithms, and sensitivity analysis, shape optimization provides a systematic approach to addressing complex real-world design problem~\cite{kanbur2020design,chen2023optimization}.

A central branch of this field is parametric shape optimization (PSO), which represents the shape of a target design using a finite set of predefined parameters~\cite{ammar2014parametric}. This method streamlines shape representations by condensing them into a manageable number of variables, thereby facilitating efficient optimization through various algorithms. Typically, these optimization algorithms are categorized into two groups: gradient-based and non-gradient-based~\cite{daxini2017parametric}.

Gradient-based methods involve computing the first-order or higher-order derivatives of performance metrics relative to shape parameters. These derivatives are utilized to guide the search direction towards an improved geometric configuration. Notable applications include using the steepest descent algorithm for optimizing hydraulic axisymmetric bodies in laminar flows~\cite{chen2021numerical}, sequential linear programming for reducing stress concentrations in shoulder fillets~\cite{pedersen1982design}, quasi-Newton methods for aerodynamic wing design~\cite{jakobsson2007mesh}, and sequential quadratic programming for shape optimization in various contexts such as solid shell structures~\cite{ramm1993shape}, cooling films~\cite{lee2010shape}, and solar reactors~\cite{tang2022inverse}.

Alternatively, non-gradient-based heuristic algorithms also have shown significant potential for optimization~\cite{munk2015topology}. 
They typically employ populations of candidate solutions to globally explore the solution space, iteratively evolving them through nature-inspired processes.
As arguably the most common heuristic method, genetics algorithm (GA) mimics the evolutionary process of species and has proven effective in various shape optimization tasks. 
Applications of GAs 
span from optimizing the structural integrity of spanners and flanges for enhanced stress distribution~\cite{woon2001structural}, to improving the acoustic performance of sound absorbers~\cite{chang2005shape}, and advancing the design of robotic arms for increased maneuverability~\cite{hsiao2020shape}.
Other notable heuristic methods include simulated annealing~\cite{sonmez2007shape}, ant colony optimization~\cite{kumar2011multi}, bat algorithm~\cite{yang2012bat}, sequential harmony search~\cite{gholizadeh2013shape}, and imperialist competitive algorithm~\cite{khalilnejad2018optimal}, among many others.

With the advent and subsequent popularity of machine learning (ML) algorithms, their applications in shape optimization have garnered significant attention~\cite{li2022machine}. Among ML paradigms, supervised learning, trained on labeled datasets, has been extensively utilized to develop surrogate models that accelerate shape optimization at substantially reduced costs. Examples include minimizing wind effects on civil structures~\cite{ding2018multi}, reducing hydraulic drag on symmetric bodies~\cite{chen2021numerical}, and enhancing the energy efficiency of bidirectional impulse turbines~\cite{ezhilsabareesh2018shape}. Leveraging inherent data structures without requiring explicit labels, unsupervised learning facilitates the reduced-order modeling of shape-induced physical fields~\cite{legresley2000airfoil,xiao2010model} or the shape itself~\cite{ghoman2012pod}, thereby enhancing optimization efficiency. Additionally, reinforcement learning (RL), a popular ML paradigm known for applications in robotic control~\cite{kober2013reinforcement} and gaming~\cite{silver2017mastering}, has also been applied to shape optimization. For instance,\cite{yan2019aerodynamic} reports on optimizing missile control surfaces using RL, and subsequent research has extended RL applications to airfoil design~\cite{viquerat2021direct,dussauge2023reinforcement}, wind-sensitive building optimization~\cite{yan2019aerodynamic}, and heat exchanger optimization~\cite{keramati2022deep}.

In-context learning (ICL) has recently emerged as a new ML paradigm, arising as an emergent capability of large language models (LLMs) when both the scale of their training datasets and model architectures surpass certain thresholds~\cite{wei2022emergent}.
Unlike traditional paradigms, 
ICL allows LLMs to perform tasks based solely on provided contextual information, without the need for explicit retraining~\cite{dong2022survey}. 
Consequently, LLMs have already shown great potential in various domains such as robotics~\cite{wang2023prompt,xu2024training}, 
industrial control~\cite{song2023pre}, and 
general-purpose optimization~\cite{guo2023towards,yang2024large,zhang2023using}. 
Inspired by the success of LLMs as decision-makers in these applications, we are motivated to explore their potential in PSO---a methodology we propose and investigate in this work, referred to as \al.
Specifically, our study focuses on two PSO problems involving fluid motion and dynamics, although the approach itself is readily applicable to other fields beyond fluid dynamics.

\section{Related works}

\subsection{LLM for mechanics}
LLMs open up new possibilities for addressing domain-specific challenges, including those in fluid mechanics and mechanics more broadly. For instance, \cite{buehler2024mechgpt} introduced MechGPT, a foundational model for mechanical and material science that can articulate related knowledge using natural language inputs. Similarly, \cite{kumar2023mycrunchgpt} proposed a ChatGPT-assisted framework that enables the use of natural language descriptions to guide research tasks exemplified by airfoil shape optimization and physics-informed neural network training. Furthermore, \citeauthor{kim2024chatgpt} \cite{kim2024chatgpt} prompted ChatGPT-4 to generate MATLAB code for solving a two-dimensional seepage flow problem using a finite-difference method. Extending the scope further, MetaOpenFOAM~\cite{chena2024metaopenfoam} leveraged multi-agent LLM collaboration to streamline computational fluid dynamics tasks, covering mesh preprocessing, simulation, and post-processing. 

In addition to serving as assistants to existing tools, LLMs can be directly utilized for scientific exploration, such as performing symbolic regression to derive the Navier-Stokes equations~\cite{du2024large} and predicting flow fields by leveraging their temporal autoregressive capabilities~\cite{zhu2024fluid}. Collectively, these research efforts  highlight the burgeoning applicability of LLMs in fluid mechanics and general mechanical applications.

\subsection{LLM for general optimization}

LLMs excel at decision-making by interpreting textual input, thereby supporting the solution of general optimization problems. For example, \citeauthor{guo2023towards}~\cite{guo2023towards} directly prompted LLMs to mimic classic optimization algorithms such as gradient descent and hill climbing for the minimizing the values of  least squares functions. 
Similarly, Google DeepMind~\cite{yang2024large} iteratively prompted LLMs to generate new solutions based on previous solutions and evaluations at each optimization step, achieving optimal parameter estimation for linear regression and determining the shortest path in the traveling salesman problem. Similar applications include LLM-based hyperparameter tuning~\cite{zhang2023using} and supply chain optimization~\cite{li2023large}. Furthermore, \citeauthor{liu2024large}~\cite{liu2024large} combined LLMs with GAs, employing LLMs as the evolutionary operator, which demonstrated better performance compared to \cite{yang2024large} in traveling salesman problems. Various approaches to leveraging LLMs to enhance classical optimization algorithms, such as evolutionary strategies~\cite{brahmachary2024large,lange2024large}, Bayesian optimization~\cite{liu2024BO}, and guided local search~\cite{ye2024reevo}, have also been explored. Moreover, LLMs have been employed to generate and design new optimization algorithms~\cite{wu2024evolutionary,zhong2024leveraging,pluhacek2023leveraging,liu2023algorithm}. However, these applications primarily address general mathematical optimization problems and have not yet explored the potential of LLMs for shape optimization or broader engineering design.

\begin{figure*}[tbh!]
\centering
\includegraphics[width=1\linewidth]{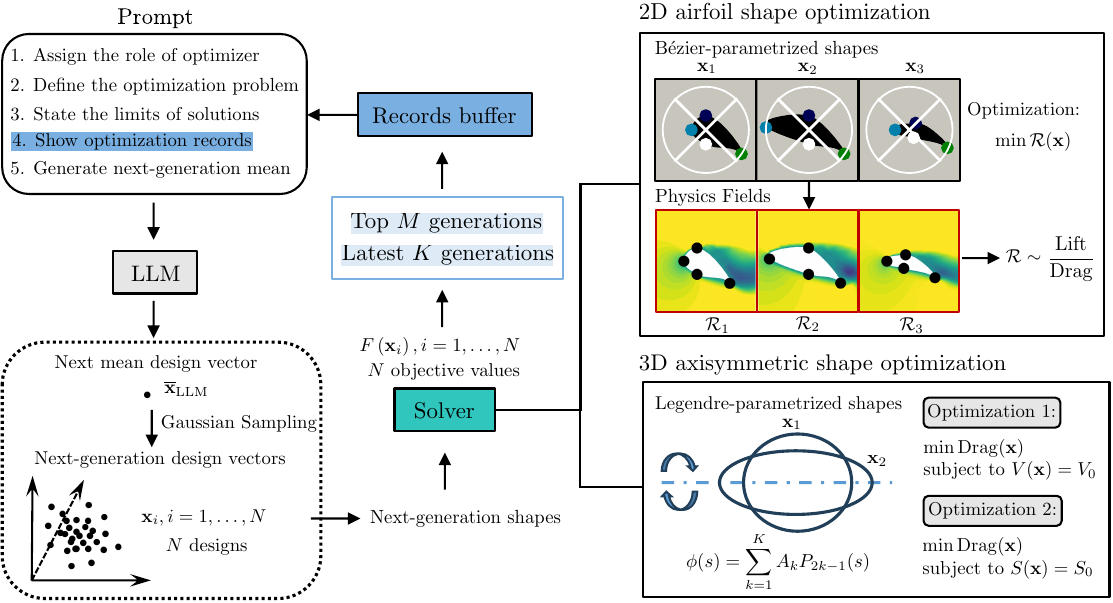}
\caption{
Overview of \al ~and its application to two PSO problems involving fluid dynamics.
}
\label{fig:wholeflowchat}
\end{figure*}

\subsection{Generative AI for engineering design}
As a broader category encompassing LLMs, generative artificial intelligence (AI) has already harnesses its powerful generative capacity for optimizing engineering design~\cite{yuksel2023review}. By acting as a robust shape generator, generative AI has demonstrated its potential to  improve design efficiency and foster innovation. Common generative AI models used in engineering design include generative adversarial networks (GANs) and diffusion models~\cite{liao2024generative}. For instance, GANs have been applied to aerodynamic optimization~\cite{li2020efficient,wang2023airfoil,usama2024generative} and building layout design \cite{jiang2023building}. More recently, diffusion models, notably Shap-E~\cite{jun2023shap}, have supported applications like automated vehicle design via natural language description~\cite{rios2023large,wong2024prompt}.
Similar research includes diffusion-based models for the mechanical structures optimization~\cite{giannone2023aligning},  building interior design~\cite{chen2023generating}, and  airfoil shape optimization~\cite{wei2024diffairfoil}.

\subsection{Attempting LLMs for shape optimization}
In summary, LLMs serve as decision-makers for addressing general-purpose optimization problems. Concurrently, their siblings---GANs and diffusion models---have been adopted in engineering design, invoking us to contempt whether LLMs possess similar potential. Hence, we attempt here exploring LLMs' decision-making abilities for a specific design task---parametric shape optimization.

\section{METHODOLOGY}

Inspired by \cite{lange2024large} that employs LLMs for optimization in the spirit of evolutionary strategy,
we introduce an LLM-based optimization framework named \al. This framework is developed to address parametric shape optimization problems. For our implementation,
we use the LLM ``claude-3-5-sonnet-20240620'' developed by Anthropic.
Before delving into the details of \al, we first define the optimization problem. A particular design is geometrically parametrized by a vector $\bx$, which encodes all optimizable design variables.  
Using \al, we seek the optimal design
\begin{align}
\bxopt=\underset{\bx}{\operatorname{\text{argmax}}} \; F (\bx),    
\end{align}
which maximizes the objective function $F(\bx)$ indicating the design performance.

\subsection{Workflow of \al}

In the vein of evolutionary strategies~\cite{li2020evolution}, \al ~functions by evolving a population of $N$ designs $\{\mathbf{x}_i\}_{i=1}^{N}$ generation by generation. Its workflow is demonstrated in Fig.~\ref{fig:wholeflowchat}. 
For simplicity, we assume that the design vectors within one generation follows an $N$-dimensional
Gaussian distribution  
with a mean  $\bar{\mathbf{x}}$ and fixed variance matrix $\sigma^2 \mathbf{I}$, \ie
 \begin{equation}
\mathbf{x} \sim \mathcal{N}(\bar{\mathbf{x}}, \sigma^2 \mathbf{I}).
 \end{equation}

We start the optimization by initializing the first $\nini$ generations of design vectors.
These generations are sampled from
the afore-mentioned Gaussian distribution, with its mean $\bar{\mathbf{x}}$ seeded randomly within a  range specific to case. 
Then we evaluate the objective function $F(\bx)$, \ie the performance  for each initial design $\bx$, and store the design-performance pair, $\left[ \bx, F(\bx) \right]$, of all such designs in a record buffer.

Subsequently, LLM guides the searching of the optimal design. 
First, we prepare a prompt (see Sec.~\ref{sec:prompt2llm}) that encompasses selected design-performance records from the buffer; the selection strategy is detailed in Sec.~\ref{sec:select}. 
The prompt demonstrates these selections to the LLM, which returns a mean $\bar{\bx}_{\rm LLM}$ for the next generation.
Then, resampling from the Gaussian distribution with this mean $\bar{\bx}_{\rm LLM}$ yields a new generation of designs.
Next, we assess these designs' performance and append the paired design-performance data to the record buffer. 
The augmented buffer contributes to the prompt preparation for the following generation, thus beginning a new iteration. 

\subsubsection{Overview of the prompt}\label{sec:prompt2llm}
\begin{figure}[tbh!]
\centering
\includegraphics[width=1\linewidth]{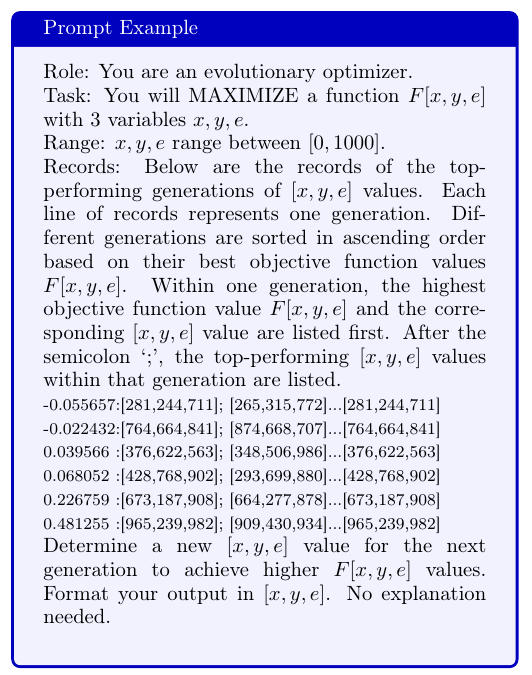}
\caption{
A specific example prompt of \al.
}
\label{fig:prompt}
\end{figure}

We develop for \al ~a few-shot prompting architecture consisting of five parts: 
\begin{enumerate}
    \item  Assign the task of evolutionary optimization to the LLM;
    \item Specify the dimensions of the design vector and the optimization objective;
    \item Delineate the parameter range for optimization;
    \item Supply the LLM with selected records and direct it to suggest the most promising mean $\bar{\bx}_{\rm LLM}$ for the subsequent generation;
    \item Instruct the LLM to present this mean $\bar{\bx}_{\rm LLM}$ in a specified format.
\end{enumerate}
An example prompt is shown in Fig.~\ref{fig:prompt}.

\subsubsection{Strategy of records selection}\label{sec:select}

Design-performance records are selectively included in the fourth part of the prompt, expressed in natural language. The selection process is as follows.
Firstly, we sort the designs within each generation in ascending order based on their objective functions $F(\mathbf{x})$. Next, we rank the generations in ascending order according to the best design (\ie the one with the highest $F(\mathbf{x})$) in each generation.
We then identify the top-ranking $T$ generations and the most recent $R$ generations, removing overlaps. 
For each of these identified generations, only the top-ranking $M$ designs are selected to be included in the prompt.

\subsection{Other details}
Due to the limitations of LLMs in handling floating-point numbers effectively, we transform the values within the design vectors into integers~\cite{xu2024training,lange2024large} ranged from $0$ to $1000$ before querying the LLM. 
To further ensure the stability of the outputs, we set the temperature parameter to $0$, thereby forcing the LLM to generate deterministic results by consistently selecting the highest-probability output.
\section{Results and observations}

\begin{figure*}[tbh!]
\centering
\includegraphics[width=1\linewidth]{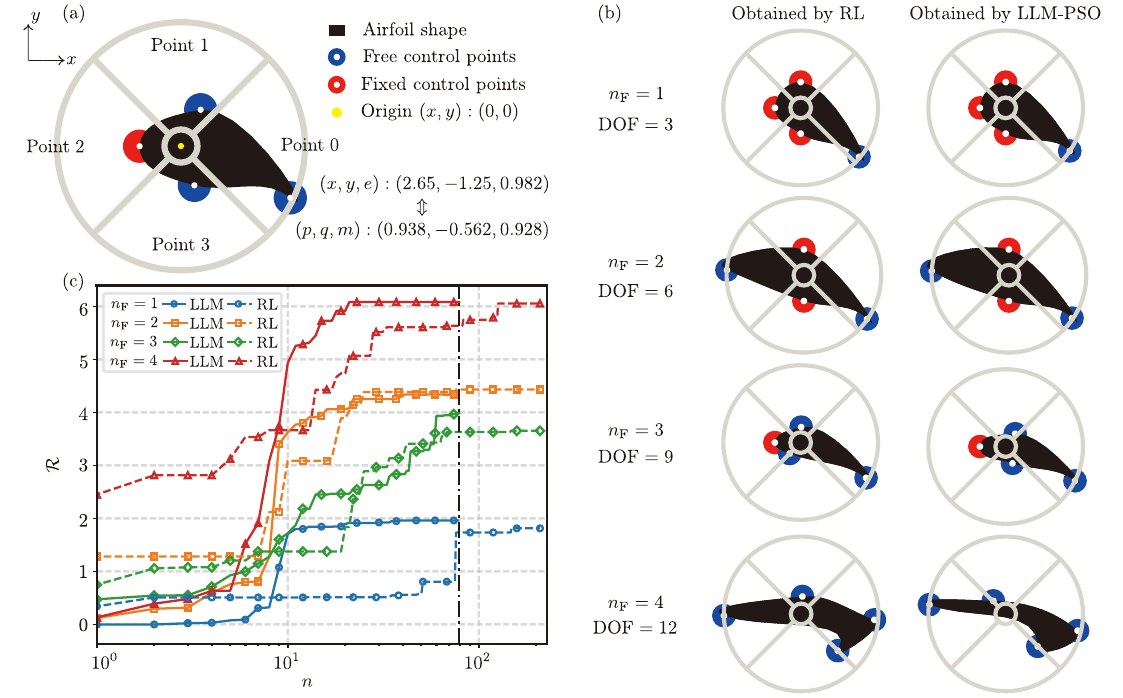}
\caption{
(a) Geometric parametrization of an airfoil profile by a four-point B\'ezier curve. 
(b) Optimal airfoils based on different number $\nfree \in [1,4]$ of free points. 
(c) Comparison of the optimization trajectories: the lift-drag ratio versus iteration number $\mathcal{R}(n)$, between \al ~and an RL algorithm~\cite{viquerat2021direct}. 
}
\label{fig:4ptsresults}
\end{figure*}

We employ \al ~to tackle two representative optimization tasks in fluid dynamics and benchmark our results against those obtained using other algorithms. Specifically, we seek the optimal shapes of a two-dimensional (2D) airfoil and a three-dimensional (3D) axisymmetric body in fluid flow.

\subsection{Two-dimensional airfoil shape optimization}
We apply our strategy to a classical setting---optimizing the shape of a 2D airfoil to maximize its aerodynamic performance at a moderate Reynolds ($\Re$) number. 
This problem~\cite{viquerat2021direct} was previously addressed using RL. In our study, we adopt the same setting but replace the RL-based optimizer with our LLM-based version, \al.

\subsubsection{Shape parametrization}
We parameterize the airfoil profile by a B\'ezier curve connecting four control points indexed $i\in\{0, 1, 2, 3\}$.
The $i$-th point is characterized by its Cartesian coordinates $(x_i,y_i)$ and a sharpness factor $e_i$ (see \SI) of the B\'ezier curve at this point. Here, the coordinates have been non-dimensionalized by a characteristic length $\ell$; the same applies to other coordinates or length scales from hereinafter. 

We then introduce a polar coordinate system $\lp \rho,\theta \rp$ equally partitioned into four sectors in analogous to a pie chart, see Fig.~\ref{fig:4ptsresults}(a). Each control point, for example the $i$-th with coordinates $\lp \rho_i,\theta_i \rp$, is restricted exclusively to one distinct sector,
namely, $\theta_i \in \left[-\frac{\pi}{8} + \frac{\pi}{4}i, \frac{\pi}{8} + \frac{\pi}{4}i\right]$. 
This restriction is imposed to reduce the occurrence of entangled airfoil profiles. Here, $\rho_i=\sqrt{x_i^2+y_i^2}$ and $\lp x_i,y_i\rp=\rho_i\lp \cos\theta_i, \sin\theta_i \rp$. Further, $(\rho_i, \theta_i, e_i)$ is transformed to 
a three-element vector 
$(p_i, q_i, m_i)\in [-1,1]^3$, 
\begin{subequations}
    \begin{align}
p_i &= 2\left(\frac{\rho_i - \rho_{\min}}{\rho_{\max}-\rho_{\min}}\right) - 1, \\
q_i &= 2\left(\frac{4}{\pi}\theta_i-i\right), \\
m_i &= 2e_i-1,        
    \end{align}
\end{subequations}
where \(\rho_{\min} = 0.6\) and \(\rho_{\max} = 3\) set the size of airfoils.

\begin{figure*}[tbh!]
\centering
\includegraphics[width=0.736842\linewidth]{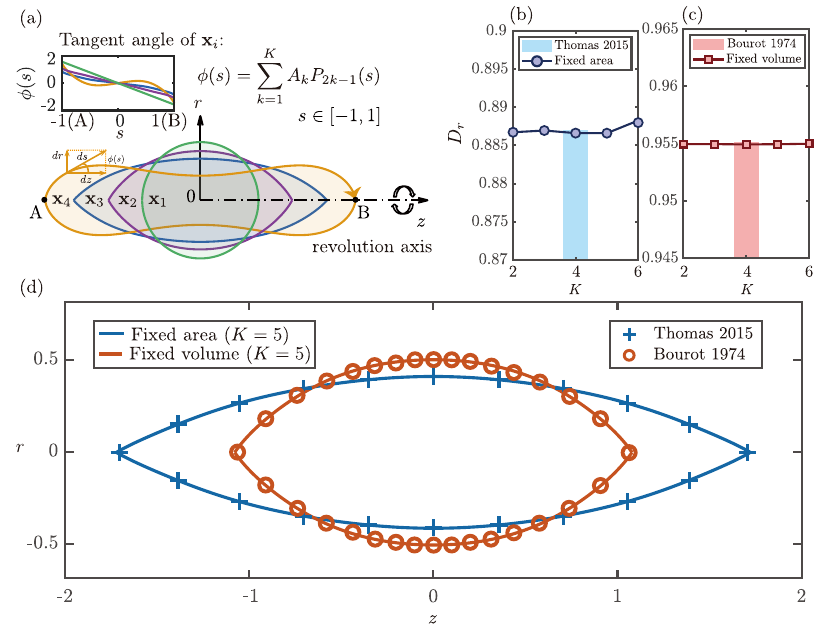}
\caption{ (a) 
Parametrization of the 2D profile of an axisymmetric body using Legendre polynomials.
The normalized drag $\Dr$ averaged over five \al-based optimizations versus the number $K$ of DOFs,
when the (b) surface area or (c) volume of the body is fixed. The results are compared to theoretical solutions.
(d) LLM-obtained optimal profiles (solid curve) in comparison to theoretical counterparts (symbol).
}
\label{fig:3dbody}
\end{figure*}

\begin{figure*}[tbh!]
\centering
\includegraphics[width=0.796842\linewidth]{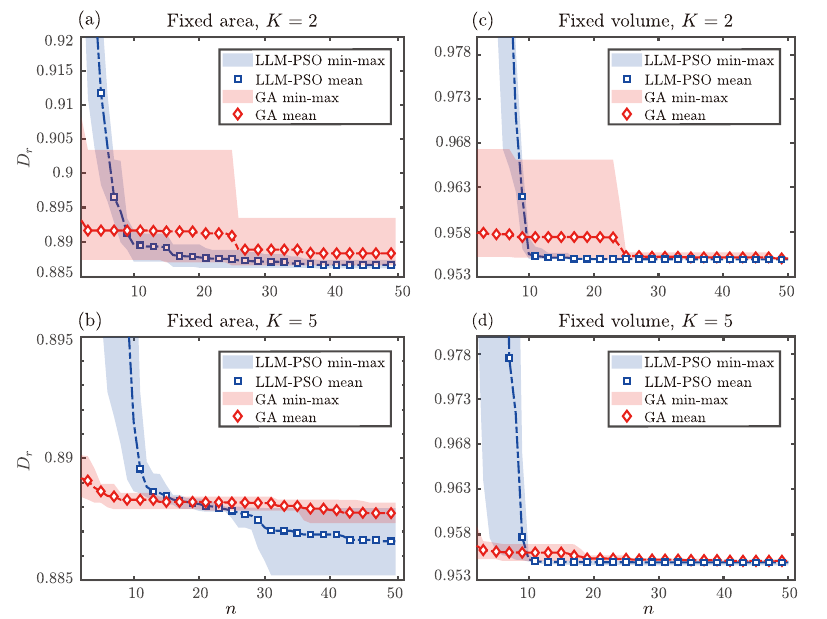}
\caption{
Comparison of the performance of \al ~and GA based on their optimization trajectories: the normalized drag versus iteration number $\Dr(n)$, represented by mean values and min-max ranges from five runs. The left and right columns present results for area-fixed and volume-fixed settings, respectively, while the top and bottom rows correspond to $K=2$ and $K=5$ degrees of freedom, respectively. To ensure a fair comparison, we adopt the same population size $N$ for \al ~and GA.
}
\label{fig:3dtrail}
\end{figure*}

\subsubsection{Optimization setup}
Using an open-source finite-element-method solver, FEniCS~\cite{alnaes2015fenics} for partial differential equations, we solve the dimensionless Navier-Stokes equation to evaluate an airfoil's aerodynamic performance (see \SI). Specifically, we focus on the ratio, $f_{\mathrm{L}}/f_{\mathrm{D}}$, of its aerodynamic lift  $f_{\mathrm{L}}$ to drag $f_{\mathrm{D}}$.
In this study, we fix the Reynolds number to $\Re=100$, with $2\ell$ serving as the characteristic length scale. At this Reynolds number, the flow past an airfoil is typically unsteady and periodic. To faithfully evaluate the performance, a simulation is first run to reach a time-periodic state. We then extract the  time-averaged lift-to-drag ratio $\langle  f_{\mathrm{L}}/|f_{\mathrm{D}}|\rangle$ as the performance indicator.

Choosing a cylinder of radius $\ell$ as the baseline design, our optimization aims to maximize the relative lift-to-drag ratio, namely,
\begin{equation}
\bar{\mR}=\left\langle\frac{f_{\mathrm{L}}}{\left|f_{\mathrm{D}}\right|}\right\rangle-\left\langle\frac{f_{\mathrm{L}}}{\left|f_{\mathrm{D}}\right|}\right\rangle_{\mathrm{cyl}}.
\end{equation}

For benchmark purposes, we adhere to \cite{viquerat2021direct} in choosing the objective function for optimization: 
\begin{equation}
{\mR}= \begin{cases}
2 \bar{\mR}, & \bar{\mR} > 0, \\
\bar{\mR}, & \bar{\mR} < 0, \\
-5, & \text{when simulation fails}.
\end{cases}
\end{equation}
Accordingly, we will seek the optimal shape parametrization $\bxopt =\underset{\bx}{\operatorname{\text{argmax}}} \; \mR (\bx)$ leading to the maximum ratio.

\subsubsection{Optimal airfoil profiles}

In the optimization, we fix certain control points and free the remaining $\nfree$ points for optimization. Because each free point has three degrees of freedom (DOFs), the total number of optimization DOFs varies from three to twelve, corresponding to $\nfree=1$ to $\nfree=4$ optimizable control points, respectively. An example of $\nfree=3$ is shown in Fig.~\ref{fig:4ptsresults}(a).

In this study, we vary $\nfree$ within the set $\left\{1,2,3,4 \right\}$ to conduct the optimization. 
The optimal shapes are illustrated in Fig.~\ref{fig:4ptsresults}(b). 
Except for $\nfree=4$, which will be discussed separately below, the obtained optimal profiles resemble the shape of classical airfoils. 
Further, we compare in Fig.~\ref{fig:4ptsresults}(c) 
the trajectories of the \al-based and RL-based optimization approaches. To ensure a fair comparison of 
their convergence, the iteration number of the latter have been scaled by the population size of the former.
Upon convergence, LLM achieves similar or slightly better optimization targets compared to RL for all cases. Notably, LLM converges faster than RL, by almost an order of magnitude when $\nfree=1$ and $4$. 
We acknowledge that when $\nfree = 4$, our \al ~cannot reproduce the optimal shape reported in \cite{viquerat2021direct}. The observation suggests that, \al ~may not perform effectively at a relatively large number of DOFs, a limitation we aim to address in future work. 
Nevertheless, it is worth-noting that the RL-based optimization does not achieve this shape either, a rarity confirmed through private communications with the authors of \cite{viquerat2021direct}.

\subsection{Drag minimization of a revolved body in Stokes flow} 
Now, we apply \al ~to seek the optimal profile of a 3D axisymmetric body for minimizing its hydrodynamic drag in Stokes flow.
This optimization problem has been addressed theoretically by fixing the  volume $V$~\cite{bourot1974numerical} or surface area $S$~\cite{Montenegro2015drag} of the body. The resulting optimal profiles can be used to validate our optimal solutions. 
Besides, we will use the classical genetic algorithm (GA) to conduct the optimization, considering that \al ~is inspired by the principles of evolutionary algorithms. 

To be general, we describe the optimization problem in the dimensionless form, choosing $\ell = \lp \frac{3V_0}{4\pi}\rp^{1/3}$ as the characteristic length scale for the case when the prescribed volume is $V_0$. 
In the alternative scenario where the surface area is fixed at $S_0$, we adopt $\ell = \lp \frac{S_0}{4\pi}\rp^{1/2}$. In both cases, $\ell$ represents the radius of a sphere with the specified volume $V_0$ or area $S_0$.

\subsubsection{Shape parametrization} 
As shown in Fig.~\ref{fig:3dbody}(a), the axisymmetric 3D body \(\Omega\) is characterized by a 2D profile in the $rz$-plane. The $z$-axis coincides with the revolution axis, and $r$ represents the radial coordinate.
This profile can be characterized by its tangent angle \(\phi(s)\) as a function of the arclength \(s \in [-1, 1]\).
Employing Legendre expansion as in \cite{Montenegro2015drag}, the angle is parameterized by 
\begin{equation}
\phi(s) = \sum_{k=1}^{K} A_k P_{2k-1}(s),
\end{equation}
which involves a number of $K$ odd Legendre polynomials 
$P_{2k-1}$ with corresponding coefficients  \(A_k\).
The coefficients \(A_1, A_2, \dots, A_K\) together constitute the $K$-dimensional design vector \(\mathbf{x}\) for this optimization case, 
\begin{equation}
\mathbf{x} = \begin{bmatrix} A_1 & A_2 & \dots & A_K \end{bmatrix}.
\end{equation}

With \(\phi(s)\) determined by $\bx$, the 2D profile $(r(s),z(s))$ can then be obtained:
\begin{equation}
r(s) = \lambda\int_{-1}^{s} \sin \phi(\ts)  \, \mathrm{d}\ts, \quad z(s) = \lambda\int_{-1}^{s} \cos \phi(\ts) \, \mathrm{d}\ts.
\end{equation}
Notably, a scaling factor $\lambda$ is needed to ensure that the obtained body of revolution satisfies the prescribed volume or area constraints.

\begin{figure}[tbh!]
\centering
\includegraphics[width=1\linewidth]{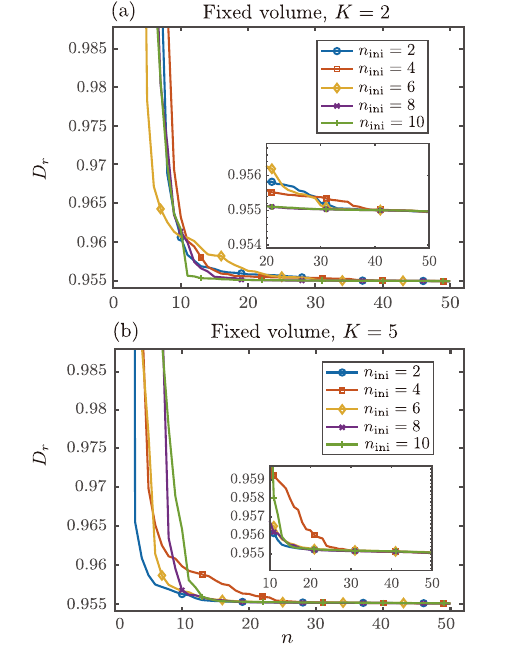}
\caption{
The influence of the number $\nini$ of randomly seeded initial generations on the optimization trajectory $\Dr(n)$ for (a) $K=2$ and (b) $K=5$. For each case, five runs are conducted to obtain a mean trajectory. 
}
\label{fig:nini}
\end{figure}

\subsubsection{Optimization setup}
We perform two types of shape optimization to minimize the drag $D$ over the body, either with a constraint on its volume $V$~\cite{bourot1974numerical} or surface area $S$~\cite{Montenegro2015drag}. 
To calculate the drag $D$, we numerically solve the axisymmetric Stokes equation using the PDE mode of COMSOL Multiphysics 5.5 (I-Math, Singapore). 
We discretize a square computational domain sized $1000 \ell$ by approximately $ 26000$ triangular Taylor-Hood elements, with local refinement near the body surface. We impose no-slip boundary condition at the body surface, a uniform velocity  at the inlet, and a zero-pressure at the outlet.

\subsubsection{Optimal profiles}
Using the algorithm and fixing either the area or the volume of the body of revolution,
we determine its drag-minimizing profiles for mode numbers $K\in [2, 6]$. Their drags, normalized by the reference drag of a sphere with the same area or volume, are denoted as $\Dr$, as presented in Fig.~\ref{fig:3dbody}(b) for the area-fixed case and Fig.~\ref{fig:3dbody}(c) for the volume-fixed case, respectively. As indicated, the minimal drags we obtain closely match with the theoretical optimal solutions~\cite{bourot1974numerical,Montenegro2015drag}. 
Naturally, the corresponding 2D profiles from our optimization, such as those at $K=5$, show excellent agreement with the benchmark profiles. These consistent results demonstrate the effectiveness of our optimization strategy.

We further compare the performance of \al  ~to that of GA in Fig.~\ref{fig:3dtrail} for $K=2$ (upper panels) and $K=-5$ (lower panels).
For comparison, we use the same population size $N$ for both algorithms and conduct five runs for each optimization task. 
In the optimization's very early stage \ie $\iter \lessapprox 5$, \al ~performs worse than GA---the drag $\Dr$ is higher in the former case. However, by $\iter \approx 10$, \al ~surpasses GA, exhibiting a steep decline in $\Dr$. Consequently, \al ~achieves a slightly better target value, \viz a lower mean drag $\Dr$ averaged over the five runs. In addition, \al ~exhibits a smaller statistical variance than GA when $K=2$. However, this trend reverses when $K$ increases to $5$.

Finally, in Fig.~\ref{fig:nini}, we examine the influence of $\nini$ on the optimization process for the volume-fixed setting. The converged minimal drag $\Dr$ is shown to be independent of $\nini$. When $K=2$, the optimization trajectories are weakly affected by $\nini$, without revealing a clear trend. As $K$ increases to $5$, we observe that a lower $\nini$ leads to faster initial convergence.

\section{Conclusions and Discussions}\label{sec:con}
We have developed a novel LLM-based 
framework \al ~that marries generative AI with evolutionary strategy for parametric shape optimization. This framework leverages the emerging ICL capacity of LLMs to assist decision-making. 
To demonstrate its effectiveness, we employed \al ~to identify drag-minimizing profiles of a 2D airfoil in laminar flow and a 3D axisymmetric body in Stokes flow. In both cases, \al ~successfully determined optimal shapes that align with established benchmark profiles.

We acknowledge that certain hyperparameters of 
\al ~have not been thoroughly investigated. For instance, the number of top-ranking designs $M$ displayed in each generation warrants additional test. Further fine-tuning these hyperparameters may enhance the optimization performance of 
\al.

Based on the current success of \al, we point several potential research directions: 1) Improving the current \al ~to address shape optimization with high DOFs;
2) Employing fine-tuning techniques to enhance the optimization capabilities of LLMs; 3) Combining other advanced optimization algorithm with LLMs to achieve synergistic enhancement in performance.

\newpage
\section*{Supplementary Information}

\subsection*{Sharpness factor in B\'ezier parametrization}

\setcounter{equation}{0}  
\renewcommand{\theequation}{A.\arabic{equation}} 

The sharpness factor plays a key role in controlling the tangent angle of the generated B\'ezier curve at a specific control point. 
Take the $i$-th point for example: We first define the relative angular orientations of point $i$ with respect to its two neighboring points as
\(\theta_{i,i-1} = \arctan{\left( \frac{y_{i} - y_{i-1}}{x_{i} - x_{i-1}} \right)}\) for point \(i\) relative to point \(i-1\) and 
\(\theta_{i,i+1} = \arctan{\left( \frac{y_{i+1} - y_{i}}{x_{i+1} - x_{i}} \right)}\) for point \(i\) relative to point \(i+1\);
Then, using these relative angles, the sharpness factor \(e_i\) determines the tangent angle \(\theta_i^{\ast}\) at
point \(i\), which is a weighted combination of the two neighboring angles:
\begin{equation}
  \theta_i^{\ast} = e_i \theta_{i-1,i} + (1 - e_i) \theta_{i,i+1}.  
\end{equation}

\subsection*{
Simulation for flow past a 2D airfoil
}
To calculate the lift-to-drag ratio of a certain airfoil design~\cite{viquerat2021direct}, we numerically solve the Naiver-Stokes equations using the open-source PDE solver FEniCs~\cite{alnaes2015fenics}. 
We adopt 
an rectangular computational domain of size $45\ell\times30\ell$, where $\ell$ represents the radius of the cylinder as the baseline design.
A no-slip boundary condition is applied at the airfoil surface, a uniform velocity condition at the inlet, and a zero-pressure condition at the outlet. 
We utilize Gmsh~\cite{geuzaine2009gmsh} to generate the mesh for this simulation.

\subsection*{Data Availability}
The datasets used and/or analyzed during the current study available from the corresponding author on reasonable request.

\subsection*{Code Availability}
The prompt along with the related source codes will be open-sourced upon the acceptance of this manuscript.

\clearpage


\end{document}